\documentclass[10pt,twocolumn,letterpaper]{article}

\usepackage[final]{cvpr}      

\usepackage{graphicx}
\usepackage{amsmath}
\usepackage{amssymb}
\usepackage{booktabs}

\usepackage{adjustbox}
\usepackage{array} 
\usepackage{xcolor,colortbl}

\usepackage[pagebackref,breaklinks,colorlinks]{hyperref}

\usepackage[capitalize]{cleveref}
\crefname{section}{Sec.}{Secs.}
\Crefname{section}{Section}{Sections}
\Crefname{table}{Table}{Tables}
\crefname{table}{Tab.}{Tabs.}

\def\cvprPaperID{4637} 
\def\confName{CVPR}
\def\confYear{2022}

\newcommand{\Amat}{{\boldsymbol A}}

\newcommand{\Dmat}{{\boldsymbol D}}

\newcommand{\Hmat}[0]{{{\boldsymbol H}}}

\newcommand{\Kmat}[0]{{{\boldsymbol K}}}

\newcommand{\Mmat}[0]{{{\boldsymbol M}}}

\newcommand{\Qmat}[0]{{{\boldsymbol Q}}}

\newcommand{\Vmat}[0]{{{\boldsymbol V}}}

\newcommand{\Xmat}{{\boldsymbol X}}
\newcommand{\Ymat}[0]{{{\boldsymbol Y}}}
\newcommand{\Zmat}{{\boldsymbol Z}}

\newcommand{\xv}{\boldsymbol{x}}
\newcommand{\yv}{\boldsymbol{y}}

\newcommand{\zv}{\boldsymbol{z}}

\newcommand{\Thetamat}{\boldsymbol{\Theta}}

\begin{document}

\title{Spectral Compressive Imaging Reconstruction Using Convolution and Contextual Transformer}

\author{%
	Lishun Wang $^{1,2}$, Zongliang Wu $^{3}$, Yong Zhong $^{1,2}$ and Xin Yuan $^3$\\
		$^{1}$ Chengdu Institute of Computer Application Chinese Academy of Sciences, \\ 
    $^2$ University of Chinese Academy of Sciences, \\
    $^3$  Westlake University
}



\maketitle

\begin{abstract}
	Spectral compressive imaging (SCI) is able to encode the high-dimensional hyperspectral image to a 2D measurement, and then uses algorithms to reconstruct the spatio-spectral data-cube. 
	At present, the main bottleneck of SCI is the reconstruction algorithm, 
	and the state-of-the-art (SOTA) reconstruction methods generally face the problem 
	of long reconstruction time and/or poor detail recovery. 
	In this paper, we propose a novel hybrid network module, namely CCoT (Convolution and Contextual Transformer) block, 
	which can acquire the inductive bias ability of convolution and the powerful modeling ability of transformer simultaneously,
	and is conducive to improving the quality of reconstruction to restore fine details. 
	We integrate the proposed CCoT block into deep unfolding framework based on the generalized alternating projection algorithm, 
	and further propose the GAP-CCoT network. 
	Finally, we apply the GAP-CCoT algorithm to SCI reconstruction. 
	Through the experiments of  extensive synthetic and real data, 
	our proposed model achieves higher reconstruction quality ($>$2dB in PSNR on simulated benchmark datasets) 
  and shorter running time than existing SOTA algorithms by a large margin. 
  The code and models are publicly available at ~\url{https://github.com/ucaswangls/GAP-CCoT}.
\end{abstract}

\section{Introduction}
\label{sec:intro}
\begin{figure}
    \centering
    \includegraphics[width=0.88\linewidth]{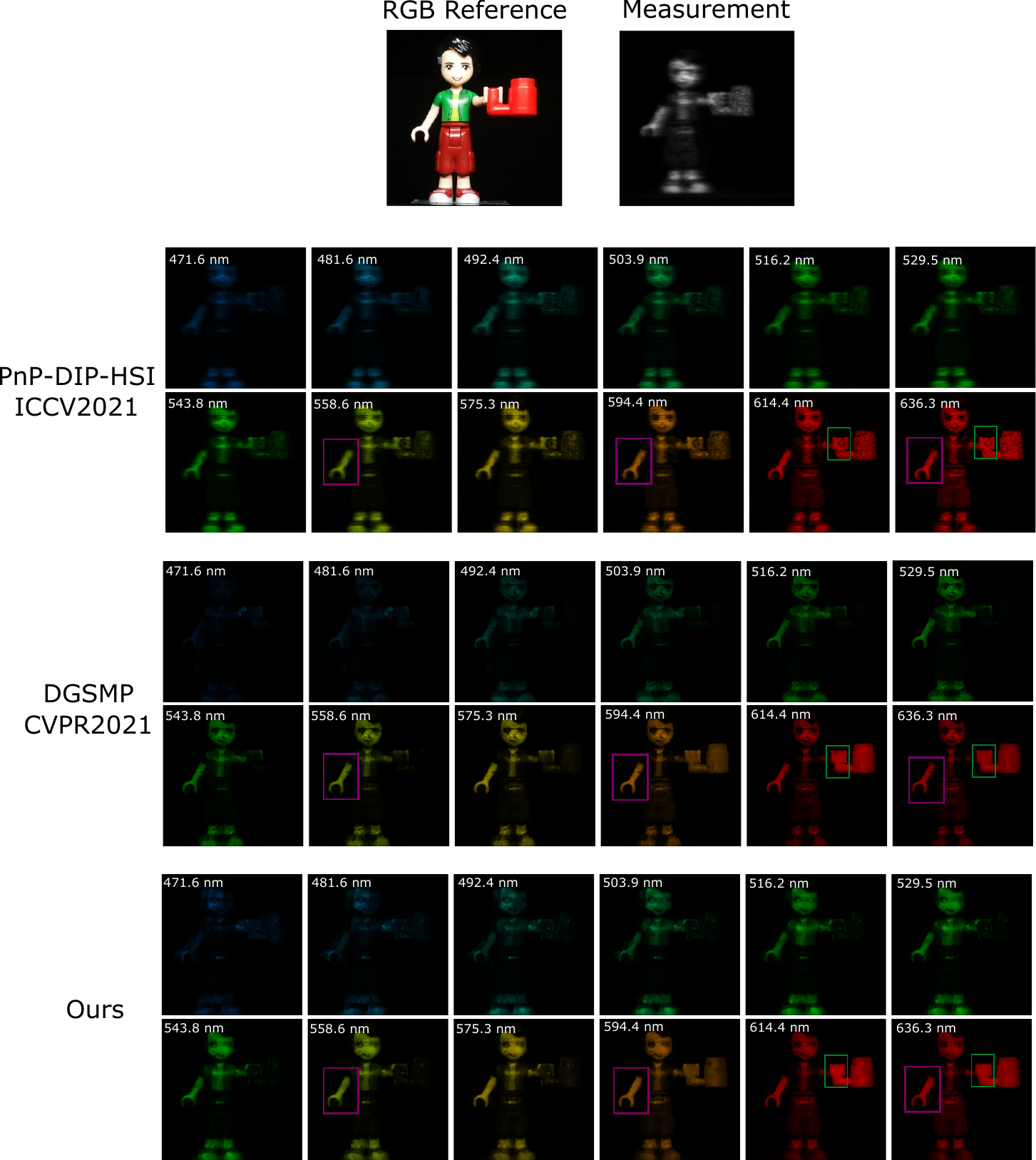}
    \caption{Reconstructed real data of \texttt{Legoman}, captured by the spectral SCI
    systems in \cite{Meng2020b}. We show the reconstruction results of 12 spectral channels,  
    and compare our proposed method with the latest self-supervised method (PnP-DIP-HSI~\cite{Meng2021b})
    and method based on the Maximum a Posterior (MAP) estimation (DGSMP algorithm~\cite{Huang2021b}).
    As can be seen from the purple and green areas in the plot, the image reconstructed by our method is clearer,
    the PnP-DIP-HSI method produced some artifacts, 
    and the DGSMP method lost some details.}
    \label{fig:best_results}
\end{figure}

Hyperspectral image is a spatio-spectral data-cube composed of many narrow spectral bands, with each one corresponding to one wavelength. 
Compared with RGB images, 
hyperspectral images have rich spectral information and can be widely used in
medical diagnosis \cite{Meng2020e}, food safety \cite{Feng2012}, remote sensing \cite{Bioucas-Dias2013}  and other fields. 
However, existing hyperspectral cameras have a long imaging time and high hardware costs, 
which greatly limits the application of these hyperspectral cameras. 
To address the above problems, spectral compressive imaging (SCI), especially the coded aperture snapshot spectral
imaging (CASSI) system \cite{Wagadarikar2008,Gehm2007,Meng2020e} provides an elegant solution, 
which can capture the information of multiple spectral bands 
at the same time with only one two-dimensional (2D) sensor.
CASSI uses a physical mask and a prism to modulate the spectral data-cube, 
and captures the {\em modulated (and thus compressed) measurement} on the 2D plane sensor.
Then reconstruction algorithms are employed to recover the hyperspectral data-cube from the measurement along with the mask.
This paper focuses on the reconstruction algorithm. 

At present, SCI reconstruction algorithms 
mainly include model-based methods and learning-based methods. 
The traditional model-based methods have relevant theoretical proof and can be explained. 
The representative algorithms are mainly TwIST \cite{Bioucas-Dias2007}, GAP-TV \cite{Yuan2016} and DeSCI \cite{Liu2018}. 
However, model-based methods require prior knowledge, long reconstruction time and usually can only provide poor reconstruction quality. 
With its strong fitting ability, deep learning model 
can directly learn relevant knowledge from data and provide excellent reconstruction results~\cite{Miao2019,Wang2021d,barbastathis2019use,fu2021coded}. 
However, compared with the model-based method, the learning-based method lacks interpretability~\cite{Yuan2021a}.

The deep unfolding network combines the advantages of model-based and learning-based methods, 
and thus it is powerful with a clear interpretability~\cite{gregor2010learning,yang2016deep,yang2018admm,zhang2018ista}.
At present, most advanced reconstruction algorithms \cite{Wang2019,Meng2020b} are based on the idea of deep unfolding.
Many models combine U-net\cite{Ronneberger2015} network with deep unfolding idea for image reconstruction and achieve good reconstruction results. 
However, the U-net model is too simple to fully obtain the effective information of the image.
Therefore, we use the inductive bias ability of convolution and 
the powerful modeling ability of transformer~\cite{han2020survey} to design a {\em parallel module} to solve the problem of SCI reconstruction. 
As shown in Fig.~\ref{fig:best_results}, the integration of our proposed method and the deep unfolding idea 
can recover more details with fewer artifacts. 
Our main contributions in this paper are summarized as follows: 
\begin{itemize} 
    \item We first apply {\bf transformer into deep unfolding} for SCI reconstruction. 
    \item We propose an effective {\bf parallel network structure composed of convolution and transformer}, dubbed CCoT, 
    which can obtain more effective spectral features. 
    \item Experimental results on a large amount of synthetic and real data 
    show that our proposed method {\bf achieves 
    state-of-the-art (SOTA) results} in the SCI reconstruction.
        \item The proposed can also {\bf be used in other compressive sensing (CS) systems}, such as video CS~\cite{Llull2013,Hitomi11_ICCV_videoCS,Reddy11_CVPR_P2C2},  
    and leads to excellent results. 
\end{itemize}

\section{Related Work}
In this section, we first review the forward model of CASSI, 
then the existing reconstruction methods are briefly introduced. 
Focusing the deep learning based models, 
we describe the pros and cons of CNN and introduce the visual transformer for other tasks. 

\subsection{Mathematical Model of SCI System}
\begin{figure}[htbp!]
  \centering
  \includegraphics[width=1.0\linewidth]{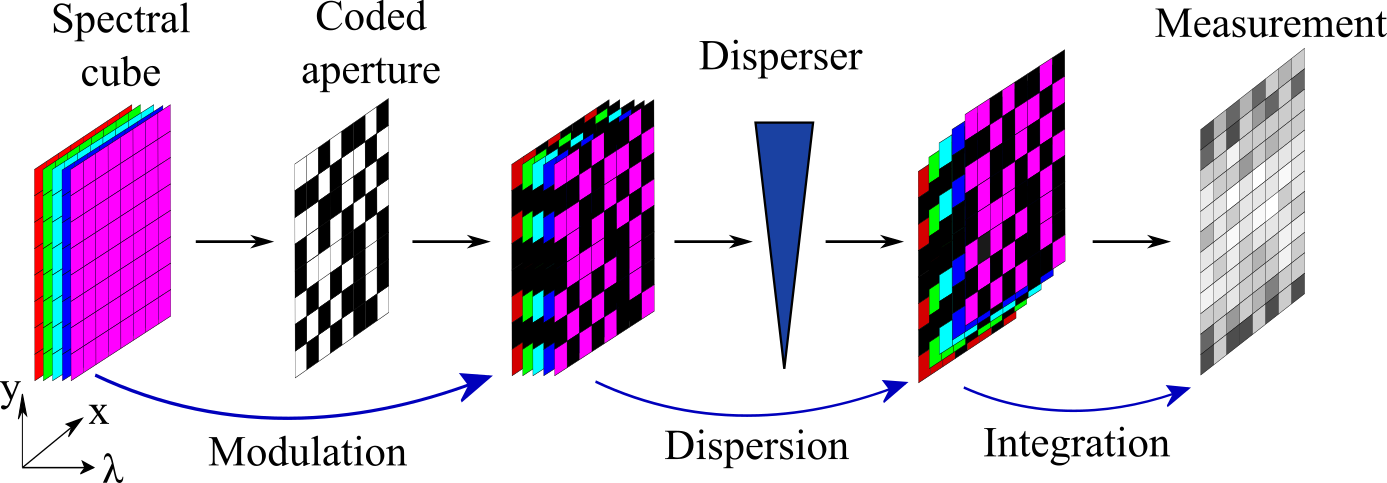}
  \caption{Schematic diagrams of CASSI system.}
  \label{fig:spectral}
  \vspace{-3mm}
\end{figure}

The SCI system encodes high-dimensional spectral data-cube into a 2D measurement, and the
CASSI \cite{Wagadarikar2008} is one of the earliest SCI systems. 
As shown in Fig.~\ref{fig:spectral}, the three-dimensional (3D) spatio-spectral data-cube is first modulated through a coded aperture (a.k.a., mask). 
Then, the encoded 3D spectral data-cube is dispersed by the dispersion prism.
Finally, the whole spectral dimension data is captured by a 2D camera sensor by integrating across the spectral dimension.

Let $\Xmat^0\in\mathbb{R}^{n_x\times{n_y}\times{n_{\lambda}}}$ denotes the captured 3D spectral data-cube, 
$\Mmat\in\mathbb{R}^{n_x\times{n_y}}$ denotes a pre-defined mask.
For each wavelength $m=1,\ldots,n_\lambda$ , the spectral image is modulated, and we can express it as
\begin{equation}
  \Xmat^{'}(:,:,m) = \Xmat(:,:,m)\odot\Mmat,  
\end{equation}
where $\Xmat^{'}\in\mathbb{R}^{n_x\times{n_y}\times{n_{\lambda}}}$
denotes the modulated spectral data-cube,
$\Xmat(:,:,m)$ denotes the $m$-th channel of the 3D spectral data-cube $\Xmat$, 
and $\odot$ denotes the element-wise multiplication. 

After passing the dispersive prism, the modulated spectral data-cube is tilted, 
and the tilted spectral data-cube $\Xmat^{''}(u,v,m)\in\mathbb{R}^{n_x\times{(n_y+n_\lambda-1)}\times{n_\lambda}}$
can be expressed as
\begin{equation}
\Xmat^{''}(u,v,m) = \Xmat^{'}(x,y+d(\lambda_m-\lambda_c),m),
\end{equation}
where $(u,v)$ represents the coordinate system of the camera detector plane,
$\lambda_m$ represents the wavelength of the $m$-th channel,
$\lambda_c$ represents the center wavelength, and $d(\lambda_m-\lambda_c)$ represents 
the spatial shifting of the $m$-th channel. 
Finally, by compressing the spectral domain, 
the camera sensor captures a 2D compressed measurement 
$\Ymat\in\mathbb{R}^{n_x\times{(n_y+n_\lambda-1)}}$, which can be expressed as
\begin{equation}
  \Ymat= \textstyle \sum_{m = 1}^{n_\lambda}{\Xmat^{''}(:,:,m)+\Zmat},
  \label{eq:Y_mat}
\end{equation}
where $\Zmat$ 
denotes the  measurement noise. 

For the sake of simple notations, as derived in~\cite{Meng2021b}, we further give the vectorized formulation expression of Eq. (\ref{eq:Y_mat}). 
Firstly, we define $\operatorname{vec}(\cdot)$ as vectorization operation of matrix.
Then we vectorize 
\begin{eqnarray}
\yv&=&\operatorname{vec}(\Ymat)\in\mathbb{R}^{n_x{(n_y+n_\lambda-1)}},\\
\zv&=&\operatorname{vec}(\Zmat)\in\mathbb{R}^{n_x{(n_y+n_\lambda-1)}},\\
\xv&=&\left[\xv_1^{\top},\ldots,\xv^{\top}_{n_{\lambda}}\right]^{\top}\in\mathbb{R}^{n_x{n_y{n_\lambda}}},
\end{eqnarray}
where $\xv_m=\operatorname{vec}(\Xmat(:,:,m)),m=1,\ldots,n_\lambda$. 
In addition, we define sensing matrix generated by coded aperture and disperser in CASSI system as
\begin{equation}
  \Hmat = \left[\Dmat_1,\ldots,\Dmat_{n_\lambda}\right]\in\mathbb{R}^{n_x{(n_y+n_\lambda-1)}\times{n_x{n_y{n_\lambda}}}}, 
  \label{eq:H}
\end{equation}
where, for $m=1,\ldots,n_\lambda$, $\Dmat_{m}=\left[\begin{array}{c}\mathbf{0}^{(1)} \\ \boldsymbol{A}_{m} \\ \mathbf{0}^{(2)}\end{array}\right] \in \mathbb{R}^{n_{x}\left(n_{y}+n_{\lambda}-1\right) \times n_{x} n_{y}}$, 
with $\Amat_m=\operatorname{Diag}(\operatorname{vec}(\Mmat)) \in \mathbb{R}^{n_{x} n_{y} \times n_{x} n_{y}}$ is a diagonal matrix and its diagonal element is $\operatorname{vec}(\Mmat)$, 
$\mathbf{0}^{(1)} \in \mathbb{R}^{(m-1)\times{n_{x}n_y}}$ and 
$\mathbf{0}^{(2)}\in \mathbb{R}^{(n_\lambda-m)\times{n_{x}n_y}}$ represent the zero matrix.
Finally, the vectorization expression of Eq. (\ref{eq:Y_mat}) is
\begin{equation}
  \yv = \Hmat{\xv}+\zv. 
  \label{eq:y_vec}
\end{equation}
After obtaining the measurement $\yv$, the next task is 
to develop a decoding algorithm. Given $\yv$ and $\Hmat$, solve $\xv$.

\subsection{Reconstruction Algorithms for SCI}
SCI reconstruction algorithms mainly focus on how to solve the ill-posed inverse problem in \eqref{eq:y_vec}, a.k.a., the reconstruction of SCI. 
Traditional methods are generally based on prior knowledge, which is used as a regularization condition to solve the problem, 
such as using total variation (TV) \cite{Bioucas-Dias2007}, sparsity \cite{Figueiredo2007}, dictionary learning \cite{Aharon2006,Yuan2015a}, non-local low rank \cite{Liu2018}, Gaussian mixture modes \cite{Yang2014} etc. 
The main problem of these algorithms is that they need to manually set prior knowledge and iteratively solve the problem, 
and the reconstruction time is long and the quality is usually not good. 

With its powerful learning capability, the neural network can directly 
learn a mapping relationship from the measurement to the original hyperspectral images, 
and the reconstruction speed can reach the millisecond level. 
End-to-end (E2E) deep learning methods (TSA-net \cite{Meng2020d}, $\lambda$-net \cite{Miao2019}, SSI-ResU-Net \cite{Wang2021d}) take the measurement and masks as inputs, 
and use only a single network to reconstruct the desired signal directly. 
Plug-and-play (PnP) methods \cite{Zheng2021a,lai2022deep} use a pre-trained network as a denoiser plugged into iterative optimization \cite{boyd2011,Yuan2016}. 
Different from PnP methods, the denoising networks in each stage of the deep unfolding methods \cite{Wang2019,Meng2020b} are independent from each other, the parameters are not shared, and can be trained end-to-end like E2E methods.

Deep unfolding has the advantages of high-speed, high quality reconstruction and also enjoys the benefits of physical-driven interpretability. Therefore, in this paper, we follow the deep unfolding framework~\cite{Meng2020b}, and propose a new deep denoiser block based on convolution and contextual transformer. The proposed module along with deep unfolding leads to SOTA results for SCI reconstruction. 

\subsection{Limitations of CNNs for Reconstruction}
Due to the local connection and shift-invariance, 
the convolutional network~\cite{lecun1995convolutional} can extract the local features of the image very well, 
and is widely used in image recognition \cite{Krizhevsky2012, He2016, Huang2017}, object detection \cite{Redmon2016}, 
semantic segmentation \cite{Long2015}, image denoising \cite{tian2020deep} and other tasks  \cite{stone2000centertrack,he2020fastreid}. 
However, its local connection property also makes it lack the ability of global perception. 
In order to improve the receptive field of convolution, 
deeper network architecture \cite{He2016} or various pooling operations \cite{Hu2018b} are often used. 
Squeeze-and-excitation network (SENet) \cite{Hu2018b} uses the channel attention mechanism~\cite{vaswani2017attention} to aggregate the global context and redistributes the weight to each channel. 
However, these methods generally lose a significant amount of detail information 
and are not friendly to image reconstruction and other tasks that need to recover local details. 

Bearing the above concerns and considering the running time, we do not use very deep network structure in our work for the SCI reconstruction, and use convolution with a sliding step size of 2 to replace the traditional max pooling operation aiming to capture the local details of the desired spatio-spectral data-cube. 

\subsection{Visual Transformers \label{Sec:ViT}}
Vision Transformer (ViT) \cite{Dosovitskiy2020} and its variants \cite{Zhu2020,Dong2021a,Touvron2021,Yuan2021b} 
have verified the effectiveness of transformer architecture in computer vision tasks. 
However, training a good ViT model requires a large number of training datasets (\ie., JFT-300M~\cite{sun2017revisiting}), 
and its computational complexity increases quadratically with the image size. 
In order to better apply transformer to computer vision related tasks, 
the latest Swin transformer \cite{Liu2021} proposes local window self attention mechanism and the shifting window method, 
which greatly reduces the computational complexity. 
The transformer network based on Swin has achieved amazing results in computer vision tasks such as image recognition \cite{Deng2009}, 
object detection \cite{Lin2014}, semantic segmentation \cite{Zhou2017, Zhou2019} and image restoration \cite{Liang2021a}, 
which further verifies the feasibility of transformer in computer vision.
In addition, most transformers, including the Swin transformer, when calculating self attention, 
all the pairwise query-key are independently learned, 
and the rich contextual relationships between them are not used. 
Moreover, the self-attention mechanism in visual transformers often ignores local feature details, 
which is not conducive to low-level image tasks such as image reconstruction.

Inspired by contextual transformer (CoT) \cite{Li2021a} and Conformer networks \cite{Peng2021a}, 
in this paper, we propose a network structure named CCoT,  
which can take advantage of convolution and transformer to extract more effective spectral features, 
and can be well applied to image reconstruction tasks such as SCI.

\begin{figure*}
    \centering
      \centerline{\includegraphics[width=1.0\textwidth]{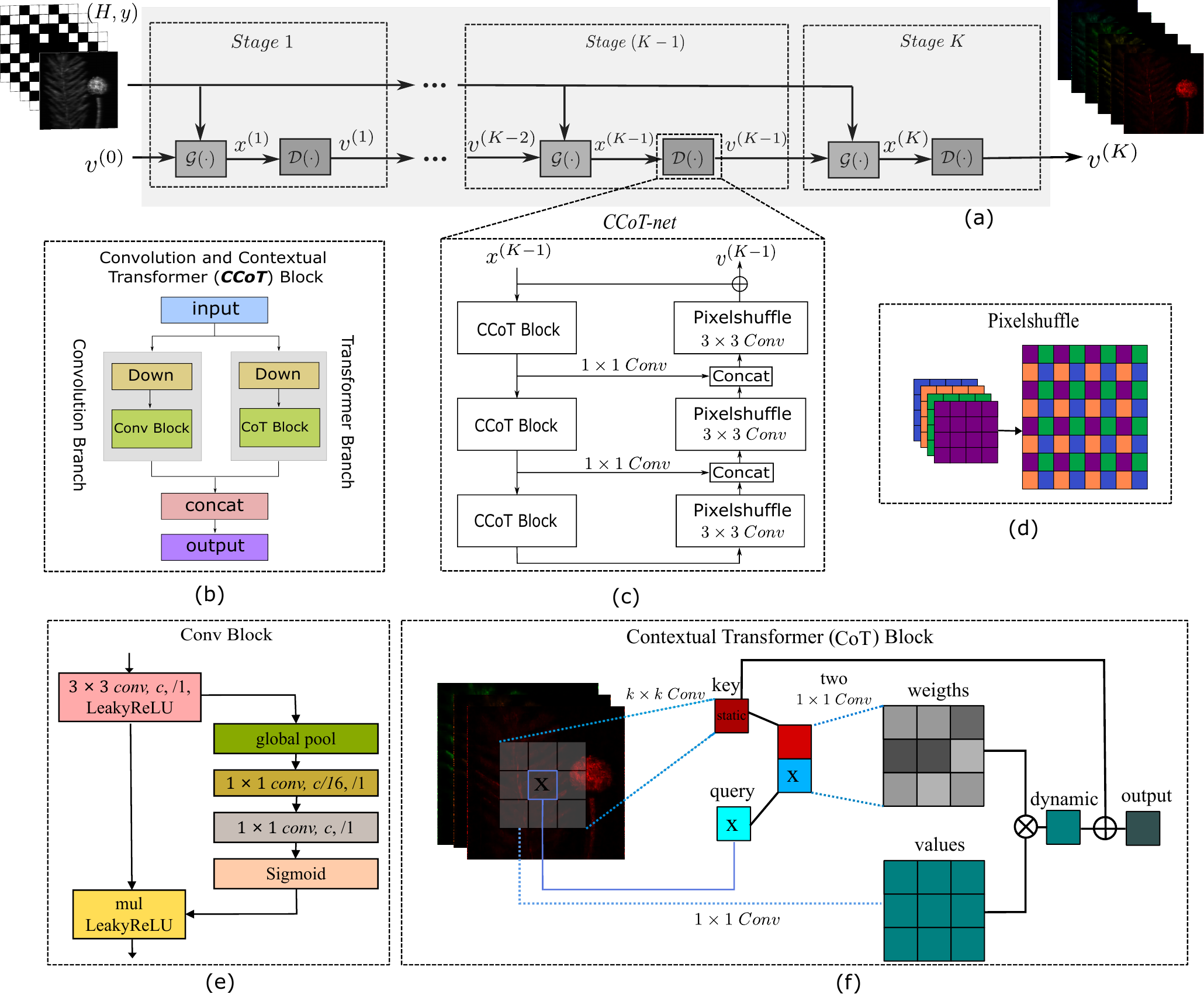}}
    \caption{Architecture of the proposed GAP-CCoT. (a) GAP-net with $K$ stages, 
    $\mathcal{G}(\cdot)$ represents the operation of Eq. (\ref{eq:iter_x}), $\mathcal{D}(\cdot)$ represents a denoiser and $\boldsymbol{v}^{(0)}=\Hmat^{\top}\yv$. 
    (b) Convolution branch and Transformer branch, The output is connected with concatenate. (c) CCoT-net, a denoiser of GAP algorithm. 
    (d) Pixelshuffle algorithm for fast upsampling. (e) Convolution block with channel attention. (f) Contextual transformer block.}
    \label{fig:network}
\end{figure*}

\section{Proposed Network}
In this section, we first briefly review the GAP-net \cite{Meng2020b} algorithm, which uses
deep unfolding ideas \cite{Hershey2014} and generalized alternating projection (GAP) algorithm \cite{Liao2014} 
for SCI reconstruction.
We select GAP-net due to its high performance, robustness and flexibility for different SCI systems reported in \cite{Meng2020b}.
Following this, we combine the advantages of convolution and transformer and then 
propose a module named convolution and contextual transformer, dubbed CCoT. 
We integrate this module into GAP-net to reconstruct hyperspectral images from the compressed measurement and masks.


\subsection{Review of GAP-net for SCI Reconstruction}
The SCI reconstruction algorithm is used to solve the following optimization problem: 
\begin{equation}
  \hat{\xv}  = \textstyle \mathop{\arg\min}\limits_{\xv}\frac{1}{2}\Vert \yv-\Hmat\xv\Vert^2+\lambda\Omega(\xv),
  \label{eq:min_x}
\end{equation}
where the first term is the fidelity term and the second term, $\Omega(\xv)$, is the prior or regularization to confine the solutions. In GAP-net and other deep unfolding algorithms, implicit priors (represented by deep neural networks) have been used to improve the performance. 

Following the framework of GAP, Eq. (\ref{eq:min_x}) can be rewritten as 
a constrained optimization problem by introducing an auxiliary parameter $\boldsymbol{v}$:
\begin{equation}
  (\hat{\xv},\hat{\boldsymbol{v}})  = \mathop{\arg\min}\limits_{\xv,\boldsymbol{v}}\frac{1}{2}\Vert \xv-\boldsymbol{v}\Vert^2_2+\lambda\Omega(\boldsymbol{v}), \
  s.t.\ \yv = \Hmat\xv.
  \label{eq:min_x_v}
\end{equation}

In order to solve Eq. (\ref{eq:min_x_v}), GAP decomposes it into the following subproblems 
for iterative solutions, where $k$ denotes the iteration number.
\begin{itemize}
\item Solving $\xv$: $\xv^{(k+1)}$ is updated via an Euclidean projection 
of $\boldsymbol{v}^{(k)}$ on the linear manifold $\mathcal{M}: \yv = \Hmat{\xv}$: 
\begin{equation}
\xv^{k+1} = \boldsymbol{v}^{(k)}+\Hmat^\top(\Hmat\Hmat^{\top})^{-1}(\yv-\Hmat{\boldsymbol{v}^{(k)}}).
\label{eq:iter_x}
\end{equation}
\item Solving $\boldsymbol{v}$: we can apply a trained denoiser to map $\xv$ closer to the desired signal space: 
\begin{equation}
\boldsymbol{v}^{k+1} = \mathcal{D}_{k+1}(\xv^{(k+1)}),   \label{eq:iter_v}
\end{equation}
where $\mathcal{D}_{k+1}(~)$ denotes the denoising operation.
\end{itemize}
It has been derived in the literature~\cite{Yuan2016} that Eq.~\eqref{eq:iter_x} has a closed-form solution due to the special structure of $\Hmat$ in Eq.~\eqref{eq:H}. Therefore, the only difference and also the novelty is the denoising step in Eq.~\eqref{eq:iter_v}. 
In the following, we describe the novel CCoT block proposed in this work for efficient and effective SCI reconstruction. 
The general reconstruction framework is illustrated in Fig.~\ref{fig:network} (a) and the detailed CCoT block is depicted in Fig.~\ref{fig:network} (b-f).

\subsection{Proposed CCoT Block for Deep Denoising}
As mentioned in Section~\ref{Sec:ViT}, to address the challenge of SCI reconstruction,
we develop the CCoT block,
where the convolution and transformer are used in parallel and can be well applied to image reconstruction tasks such as SCI.

\noindent  {\bf Convolution Branch}. As shown in Fig.~\ref{fig:network} (b,e), 
the convolution branch consists of a down-sampling layer and a channel attention (CA) block. 
In this paper, we use convolution layer to perform down-sampling by sliding step $s$ instead of the direct max pooling to capture fine details.
The channel attention block draws lessons from the idea of SENet network \cite{Hu2018b}, 
to automatically obtain the importance of each feature channel through learning,
and then to improve the useful features according to this importance 
and suppress the features that are not significant for the current task. 
The first convolution layer and channel attention module are followed by a LeakyReLU activation function \cite{Xu2015}.
The proposed convolution branch can extract local features of image well.

\begin{table*}[htp!]
    \renewcommand{\arraystretch}{1.0}
    \caption{The average PSNR in  dB (upper entry in each cell) and SSIM (lower entry in each cell)
     of different algorithms on 10 synthetic datasets. Best results are in bold.}
     \vspace{-3mm}
    \newcommand{\tabincell}[2]{\begin{tabular}{@{}#1@{}}#2\end{tabular}}
    \centering
    \resizebox{.95\textwidth}{!}
    {
    \centering
    \begin{tabular}{c|c|c|c|c|c|c|c|c|c|c|c}
    \hline
    Algorithms 
    & Scene1
    & Scene2
    & Scene3
    & Scene4
    & Scene5
    & Scene6
    & Scene7
    & Scene8
    & Scene9
    & Scene10
    & Average
    \\
    \hline
    TwIST \cite{Bioucas-Dias2007}
    &\tabincell{c}{24.81\\0.730}
    &\tabincell{c}{19.99\\0.632}
    &\tabincell{c}{21.14\\0.764}
    &\tabincell{c}{30.30\\0.874}
    &\tabincell{c}{21.68\\0.688}
    &\tabincell{c}{22.16\\0.660}
    &\tabincell{c}{17.71\\0.694}
    &\tabincell{c}{22.39\\0.682}
    &\tabincell{c}{21.43\\0.729}
    &\tabincell{c}{22.87\\0.595}
    &\tabincell{c}{22.44\\0.703}
    \\
    \hline
    GAP-TV \cite{Yuan2016}
    &\tabincell{c}{25.13\\0.724}
    &\tabincell{c}{20.67\\0.630}
    &\tabincell{c}{23.19\\0.757}
    &\tabincell{c}{35.13\\0.870}
    &\tabincell{c}{22.31\\0.674}
    &\tabincell{c}{22.90\\0.635}
    &\tabincell{c}{17.98\\0.670}
    &\tabincell{c}{23.00\\0.624}
    &\tabincell{c}{23.36\\0.717}
    &\tabincell{c}{23.70\\0.551}
    &\tabincell{c}{23.73\\0.683}
    \\
    \hline
    DeSCI \cite{Liu2018}
    &\tabincell{c}{27.15\\0.794}
    &\tabincell{c}{22.26\\0.694}
    &\tabincell{c}{26.56\\0.877}
    &\tabincell{c}{39.00\\0.965}
    &\tabincell{c}{24.80\\0.778}
    &\tabincell{c}{23.55\\0.753}
    &\tabincell{c}{20.03\\0.772}
    &\tabincell{c}{20.29\\0.740}
    &\tabincell{c}{23.98\\0.818}
    &\tabincell{c}{25.94\\0.666}
    &\tabincell{c}{25.86\\0.785}
    \\
    \hline
     HSSP \cite{Wang2019}
    &\tabincell{c}{31.48\\0.858}
    &\tabincell{c}{31.09\\0.842}
    &\tabincell{c}{28.96\\0.832}
    &\tabincell{c}{34.56\\0.902}
    &\tabincell{c}{28.53\\0.808}
    &\tabincell{c}{30.83\\0.877}
    &\tabincell{c}{28.71\\0.824}
    &\tabincell{c}{30.09\\0.881}
    &\tabincell{c}{30.43\\0.868}
    &\tabincell{c}{28.78\\0.842}
    &\tabincell{c}{30.35\\0.852}
    \\
    \hline
    $\lambda$-net \cite{Miao2019}
    &\tabincell{c}{30.82\\0.880}
    &\tabincell{c}{26.30\\0.846}
    &\tabincell{c}{29.42\\0.916}
    &\tabincell{c}{36.27\\0.962}
    &\tabincell{c}{27.84\\0.866}
    &\tabincell{c}{30.69\\0.886}
    &\tabincell{c}{24.20\\0.875}
    &\tabincell{c}{28.86\\0.880}
    &\tabincell{c}{29.32\\0.902}
    &\tabincell{c}{27.66\\0.843}
    &\tabincell{c}{29.25\\0.886}
    \\
    \hline
    TSA-net \cite{Meng2020d}
    &\tabincell{c}{31.26\\0.887}
    &\tabincell{c}{26.88\\0.855}
    &\tabincell{c}{30.03\\0.921}
    &\tabincell{c}{39.90\\0.964}
    &\tabincell{c}{28.89\\0.878}
    &\tabincell{c}{31.30\\0.895}
    &\tabincell{c}{25.16\\0.887}
    &\tabincell{c}{29.69\\0.887}
    &\tabincell{c}{30.03\\0.903}
    &\tabincell{c}{28.32\\0.848}
    &\tabincell{c}{30.15\\0.893}
    \\
    \hline
    PnP-DIP-HSI \cite{Meng2021b}
    &\tabincell{c}{32.70\\0.898}
    &\tabincell{c}{27.27\\0.832}
    &\tabincell{c}{31.32\\0.920}
    &\tabincell{c}{40.79\\0.970}
    &\tabincell{c}{29.81\\0.903}
    &\tabincell{c}{30.41\\0.890}
    &\tabincell{c}{28.18\\0.913}
    &\tabincell{c}{29.45\\0.885}
    &\tabincell{c}{34.55\\0.932}
    &\tabincell{c}{28.52\\0.863}
    &\tabincell{c}{31.30\\0.901}
    \\
    \hline
    GAP-net \cite{Meng2020b}
    &\tabincell{c}{33.03\\0.921}
    &\tabincell{c}{29.52\\0.903}
    &\tabincell{c}{33.04\\0.940}
    &\tabincell{c}{41.59\\0.972}
    &\tabincell{c}{30.95\\0.924}
    &\tabincell{c}{32.88\\0.927}
    &\tabincell{c}{27.60\\0.921}
    &\tabincell{c}{30.17\\0.904}
    &\tabincell{c}{32.74\\0.927}
    &\tabincell{c}{29.73\\0.901}
    &\tabincell{c}{32.13\\0.924}
    \\
    \hline
    DGSMP \cite{Huang2021b}
    &\tabincell{c}{33.26\\0.915}
    &\tabincell{c}{32.09\\0.898}
    &\tabincell{c}{33.06\\0.925}
    &\tabincell{c}{40.54\\0.964}
    &\tabincell{c}{28.86\\0.882}
    &\tabincell{c}{33.08\\0.937}
    &\tabincell{c}{30.74\\0.886}
    &\tabincell{c}{31.55\\0.923}
    &\tabincell{c}{31.66\\0.911}
    &\tabincell{c}{31.44\\0.925}
    &\tabincell{c}{32.63\\0.917}
    \\
    \hline
    SSI-ResU-Net (v1) \cite{Wang2021d}
    &\tabincell{c}{34.06\\0.926}
    &\tabincell{c}{30.85\\0.902}
    &\tabincell{c}{33.14\\0.924}
    &\tabincell{c}{40.79\\0.970}
    &\tabincell{c}{31.57\\0.939}
    &\tabincell{c}{{\bf 34.99}\\0.955}
    &\tabincell{c}{27.93\\0.861}
    &\tabincell{c}{{\bf 33.24}\\0.949}
    &\tabincell{c}{33.58\\0.931}
    &\tabincell{c}{31.55\\0.934}
    &\tabincell{c}{33.17\\0.929}
    \\
    \hline
    \rowcolor{lightgray}
    Ours
    &\tabincell{c}{{\bf 35.17}\\{\bf 0.938}}
    &\tabincell{c}{{\bf 35.90}\\{\bf 0.948}}
    &\tabincell{c}{{\bf 36.91}\\{\bf 0.958}}
    &\tabincell{c}{{\bf 42.25}\\{\bf 0.977}}
    &\tabincell{c}{{\bf 32.61}\\{\bf 0.948}}
    &\tabincell{c}{34.95\\ \bf{0.957}}
    &\tabincell{c}{{\bf 33.46}\\{\bf 0.923}}
    &\tabincell{c}{33.13\\{\bf 0.952}}
    &\tabincell{c}{{\bf 35.75}\\{\bf 0.954}}
    &\tabincell{c}{\bf 32.43\\{\bf 0.941}}
    &\tabincell{c}{{\bf 35.26}\\{\bf 0.950}}
    \\
    \hline
    \end{tabular}
    }
    \label{Tab:simu}
\end{table*}

\begin{figure*}
  \centering
  \begin{subfigure}{0.49\linewidth}
    \includegraphics[width=1.0\linewidth]{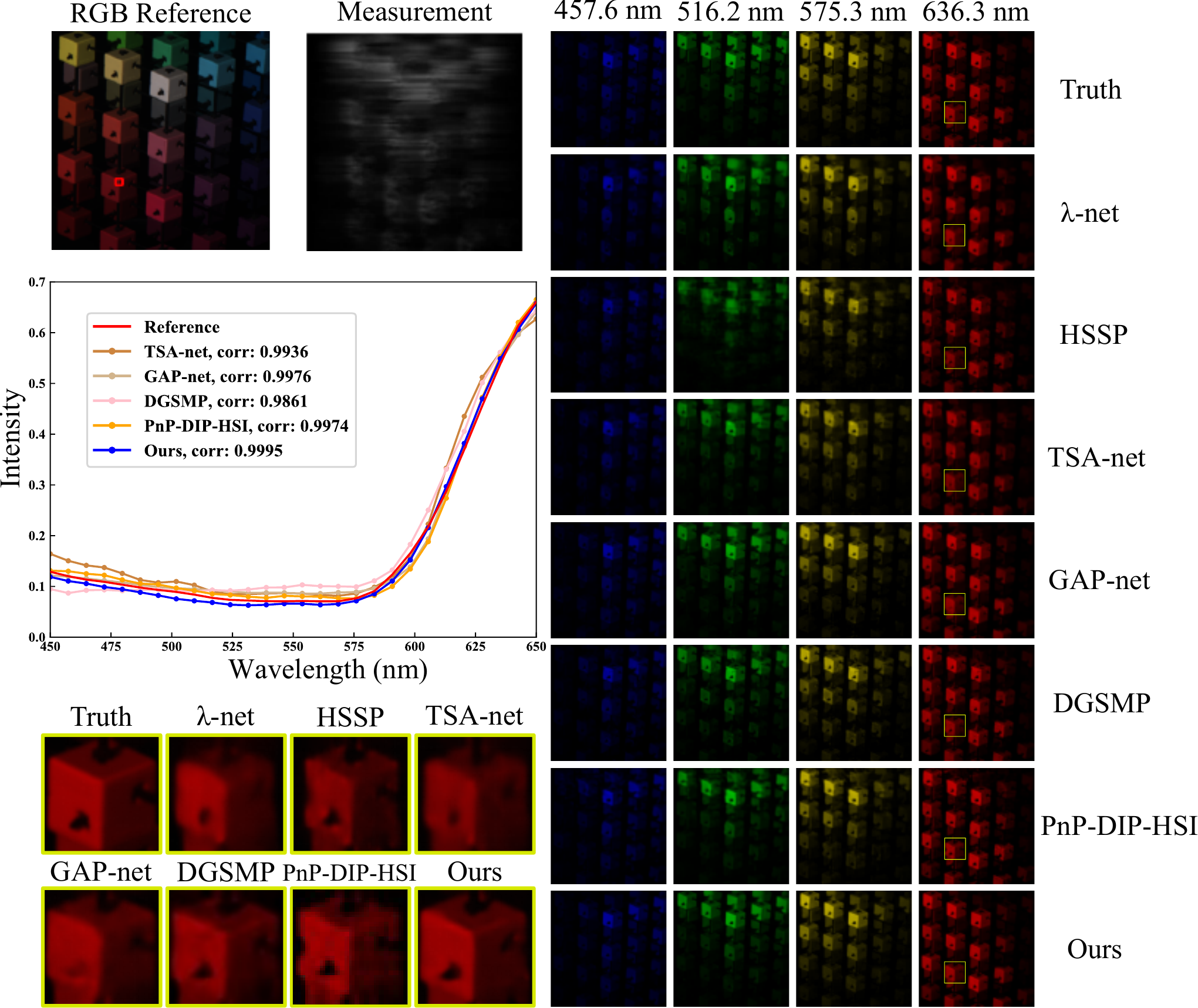}
    \label{fig:short-a}
  \end{subfigure}
  \hfill
  \begin{subfigure}{0.49\linewidth}
    \includegraphics[width=1.0\linewidth]{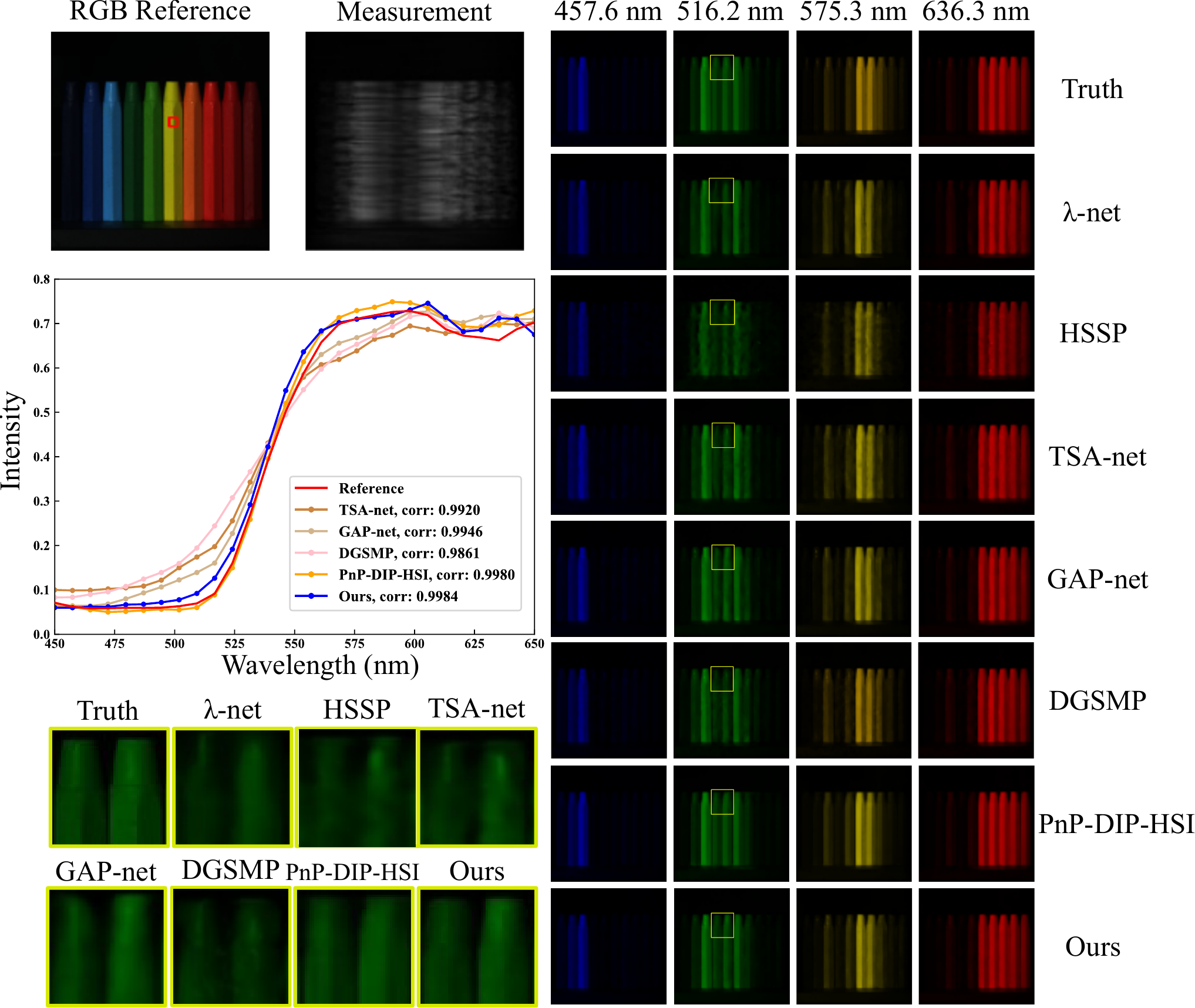}
    \label{fig:short-b}
  \end{subfigure}
  \vspace{-5mm}
  \caption{Reconstruction results of GAP-CSCoT and other spectral reconstruction algorithms 
  ($\lambda$-net, HSSP, TSA-net, GAP-net, DGSMP, PnP-DIP-HSI) in Scene 3 (left) and Scene 9 (right).
  Zoom in for better view.}
  \label{fig:sim}
  \vspace{-2mm}
\end{figure*}
  

\noindent {\bf Contextual Transformer Branch}. 
By calculating the similarity between pixels, 
the traditional transformer makes the model focus on different regions and extract more effective features. 
However, when calculating paired query-key, they are relatively independent of each other. 
A single spectral image itself contains rich contextual information, 
and there is also a significant amount of correlation between adjacent spectra. 
Therefore, we designed a contextual transformer (CoT) branch 
to better obtain features of hyperspectral images. 

As shown in Fig.~\ref{fig:network} (b), 
CoT branch consists of a down-sampling layer and a CoT block. 
The structure of the down-sampling layer is the same as the convolution branch. 
As shown in Fig.~\ref{fig:network} (f), 
we first recall that the input of the hyperspectral image is of $\Xmat^{0} \in \mathbb{R}^{n_x\times{n_y}\times{n_\lambda}}$, 
where $n_x$, $n_y$ and $n_\lambda$ represent the height, width 
and channel number of the spectral image, respectively.
{Then we define the queries, the keys, and the values as $\Kmat\in \mathbb{R}^{n_x\times{n_y}\times{n_\lambda}}$, $\Qmat \in \mathbb{R}^{n_x\times{n_y}\times{n_\lambda}}$, $\Vmat\in \mathbb{R}^{n_x\times{n_y}\times{n_\lambda}}$ respectively. }
Different from the traditional self-attention using $1\times{1}$ convolutions 
to generate mutually independent paired query-key, 
the CoT block first applies the group convolution of size $k\times{k}$
to generate a static key $\Kmat^{(1)}\in\mathbb{R}^{n_x\times{n_y}\times{n_\lambda}}$ containing the context,
and $\Kmat^{(1)}$ can be used as a static context representation of input $\Xmat^{(0)}$. $\Qmat$ and $\Vmat$ 
can be generated by the traditional self-attention mechanism. 
Then, we concatenate $\Kmat^{(1)}$ and $\Qmat$ by the 3rd dimension (spectral channels), followed by two $1\times{1}$ convolutions 
to generate an attention matrix:
\begin{equation}
  \Amat={\rm Conv}_{\delta}({\rm Conv}_\theta([\Kmat^{(1)},\Qmat]_{3})), 
\end{equation}
where $[~]_3$ denotes the concatenation along the 3rd dimension,  ${\rm Conv}_{\delta}, {\rm Conv}_{\theta}$ represent two $1\times{1}$ convolutions, $\Amat 
\in\mathbb{R}^{n_x\times{n_y}\times{(k^2\times{C_h})}}$ represents the attention matrix containing context, $C_h$ represents the number of attention heads.
{We use the traditional self-attention mechanism to perform a weighted summation 
of $\Vmat$ through $\Amat$ to obtain the dynamic context $\Kmat^{(2)} \in\mathbb{R}^{n_x\times{n_y}\times{n_\lambda}}$, and
then fuse dynamic context $\Kmat^{(2)}$ and static context $\Kmat^{(1)}$ 
as the output of the CoT block through the attention mechanism \cite{Hu2018b}.}

Finally, we concatenate the output of the convolution branch and the CoT branch as the final output of the CCoT block.

\subsection{GAP-CCoT Network}
As shown in Fig.~\ref{fig:network} (c), 
we use CCoT module and pixelshuffle algorithm to construct a U-net \cite{Ronneberger2015} like network 
as the denoiser in the GAP-net. 
The network consists of a contracting path and an expansive path. 
$i$) The contracting path contains three CCoT modules 
and $ii$) the expansive path contains three upsampling modules.
Each module of the expansive path is first quickly upsampled by the pixelshuffle algorithm \cite{Shi2016}, 
and then followed by a $3\times{3}$ convolution, 
and finally concatenates the output from the corresponding stage of the contracting path (after a $1\times{1}$ convolution) 
as the input of the next module.
Eventually, CCoT, GAP and deep unfolding  
form the reconstruction network (GAP-CCoT) of SCI. 

\noindent {\bf Loss function}. Lastly, the loss function of the proposed model is
\begin{equation}
  \mathcal{L}_{MSE}(\Theta)=\textstyle \frac{1}{n_\lambda}\sum_{n = 1}^{n_\lambda} \Vert \hat{\Xmat}_n-\Xmat_n\Vert^2,
  \label{eq:loss}
\end{equation}
where $\mathcal{L}_{MSE}(\Thetamat)$ represents the  Mean Square Error (MSE) loss,
$n_\lambda$ again represents the spectral channel to be reconstructed and $ \hat{\Xmat}_n \in {\mathbb R}^{n_x\times n_y}$ is the reconstructed hyperspectral image at the $n$-th spectral channel.

\section{Experimental Results}
In this section, we compare the performance of the proposed
GAP-CCoT network with several SOTA methods on both simulation and real datasets.
The peak-signal-to-noise-ratio (PSNR) and the the structured similarity index metrics (SSIM) \cite{Wang2004} are used 
to evaluate the performance of different HSI reconstruction methods.

\subsection{Datasets}
We use the hyperspectral dataset CAVE \cite{Yasuma2010} for model training and KAIST \cite{Choi2017} for model simulation testing. 
The CAVE dataset consists of 32 scenes, including full spectral resolution reflectance data from 400nm to 700nm with 10nm steps, 
and its spatial resolution is $512\times{512}$. 
The KAIST dataset consists of 30 scenes with a spatial resolution of $2704\times{3376}$. 
In order to match the wavelength of the real CASSI system, 
we follow the method proposed by TSA-net \cite{Meng2020d} and employ the spectral interpolation method 
to modify the training set and test data wavelength. 
The final wavelength was fitted to 28 spectral bands ranging from 450nm to 650nm.

\subsection{Implementation Details}
In the training process, we use random cropping, 
rotation, and  flipping for the CAVE dataset augmentation. 
By simulating the imaging process of CASSI,  we can obtain the corresponding measurement. 
We use measurement and mask as inputs to train the GAP-CCoT and use Adam optimizer \cite{Kingma2014}
to optimize the model. 
The learning rate is set to be 0.001 initially
and reduces by 10$\%$ every 10 epochs. Our model is trained for 200
epochs in total. 
All experiments are running on the NVIDIA RTX 8000 GPU using PyTorch. 

Finally, we use a GAP-CCoT network with 9 stages as the reconstruction network, and no noise is added to the measurement during the training process for the simulation data.
We added the shot noise to the measurements for the model training in the real data following the procedure in~\cite{Meng2020b}.

\subsection{Simulation Results}
We compared the method proposed in this paper with several SOTA methods (TwIST \cite{Bioucas-Dias2007}, 
GAP-TV \cite{Yuan2016}, DeSCI \cite{Liu2018}, 
HSSP \cite{Wang2019}, $\lambda$-net \cite{Miao2019}, TSA-net \cite{Meng2020d}, 
GAP-net \cite{Meng2020b}, PnP-DIP-HSI \cite{Meng2021b},
DGSMP \cite{Huang2021b} and SSI-ResU-Net (v1) \cite{Wang2021d}) on synthetic datasets. 
Table~\ref{Tab:simu} shows the average PSNR and SSIM results of different spectral reconstruction algorithms. 
We can see that the average PSNR value of our proposed algorithm is 35.26 dB, the average SSIM value is 0.950. 
The average PSNR value is improved by 2.09 dB than the current best algorithm SSI-ResU-Net (v1, pre-printed, not published), 
and the SSIM value is improved by 0.021. 
In addition, compared with the self supervised learning method PnP-DIP-HSI and DGSMP method (best published results)
based on the Maximum a Posterior (MAP) estimation, 
the average PSNR of the our proposed method is 3.96 dB and 2.63 dB higher, respectively. 
Based on these significant improvement, we can conduct the powerful learning capability of transformer and the proposed CCoT block. 

Fig.~\ref{fig:sim} shows part of the visualization results and spectral curves of two scenes using 
several SOTA spectral SCI reconstruction algorithms. 
Enlarging the local area, we can see that compared with other  algorithms, 
our proposed method can recover more edge details and better spectral correlation. 

\begin{table}
   \caption{The average PSNR (left entry) and SSIM (right entry) results on synthetic with
   different masks.}
    \centering
    \begin{tabular}{c|c}
    \hline
    Mask
    &PSNR, SSIM
    \\
    \hline
    Mask used in training
    &35.26, 0.950
    \\
    \hline
    New Mask1
    &35.10, 0.949
    \\
    \hline
    New Mask2
    &35.06, 0.948
    \\
    \hline
    New Mask3
    &35.06, 0.949
    \\
    \hline
    New Mask4
    &35.02, 0.948
    \\
    \hline
    New Mask5
    &34.99, 0.948
    \\
    \hline
    \end{tabular}
    \label{Tab:mask}
\end{table}

\begin{figure}
    \centering
      \centerline{\includegraphics[width=1.0\linewidth]{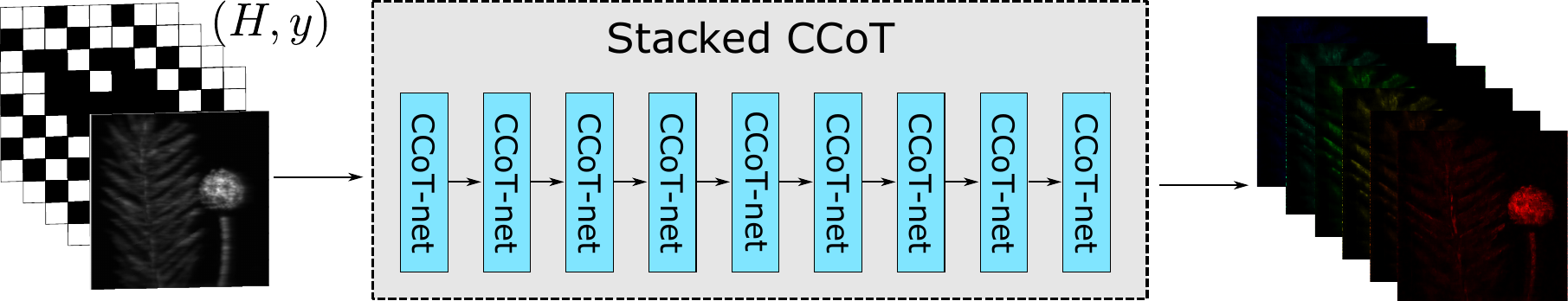}}
    \caption{Architecture of the proposed Stacked CCoT. The input of the network is $\Hmat^{\top}\yv$, the CCoT-net is the same as Fig.~\ref{fig:network} (c)}
    \label{fig:stack_cot}
    \vspace{-3mm}
\end{figure}

\subsection{Flexibility of GAP-CCoT to Mask Modulation}
The CCoT-net only serves as a denoiser for the GAP algorithm, 
so the GAP-CCoT network proposed in this paper is thus flexible for different signal modulations. 
In order to verify this point, we train GAP-CCoT network on one mask 
and test on the other five different untrained masks. 
Table~\ref{Tab:mask} shows the test results of the average PSNR value and SSIM value 
on 10 simulation data using different masks
(5 new masks of size $256\times{256}$ randomly cropped from the real mask of size $660\times{660}$). 
We can observe that for a new mask that does not appear in training, 
the average PSNR decline is maintained within 0.27 dB, 
and the result is still better than other algorithms. Therefore, 
We can conclude that the GAP-CCoT network proposed in this paper is flexible for large-scale SCI reconstruction. 

\subsection{Ablation Study}

\begin{table}[htp!]
    \renewcommand{\arraystretch}{1.0}
    \newcommand{\tabincell}[2]{\begin{tabular}{@{}#1@{}}#2\end{tabular}}
     \vspace{-2mm}
    \caption{Ablation Study: The average PSNR and SSIM values by different algorithms on 10 synthetic data.}
    \vspace{-2mm}
    \centering
    \resizebox{\linewidth}{!}
    {
    \centering
    \begin{tabular}{|c|c|c|c|c|c|c|c|c|c|c|c}
    \hline
    Algorithms 
    & \tabincell{c}{Stacked CCoT\\w/o CoT}
    & \tabincell{c}{GAP-CCoT\\w/o CoT}
    & Stacked CCoT
    & GAP-CCoT
    \\
    \hline
    PSNR/SSIM
    &32.86, 0.924
    &34.13, 0.933
    &34.27, 0.936
    &35.26, 0.950
    \\
    \hline
    \end{tabular}
    }
    \vspace{-2mm}
    \label{Tab:ablation}
\end{table}
\begin{figure}[!htbp]
    \centering 
    \includegraphics[width=.7\linewidth]{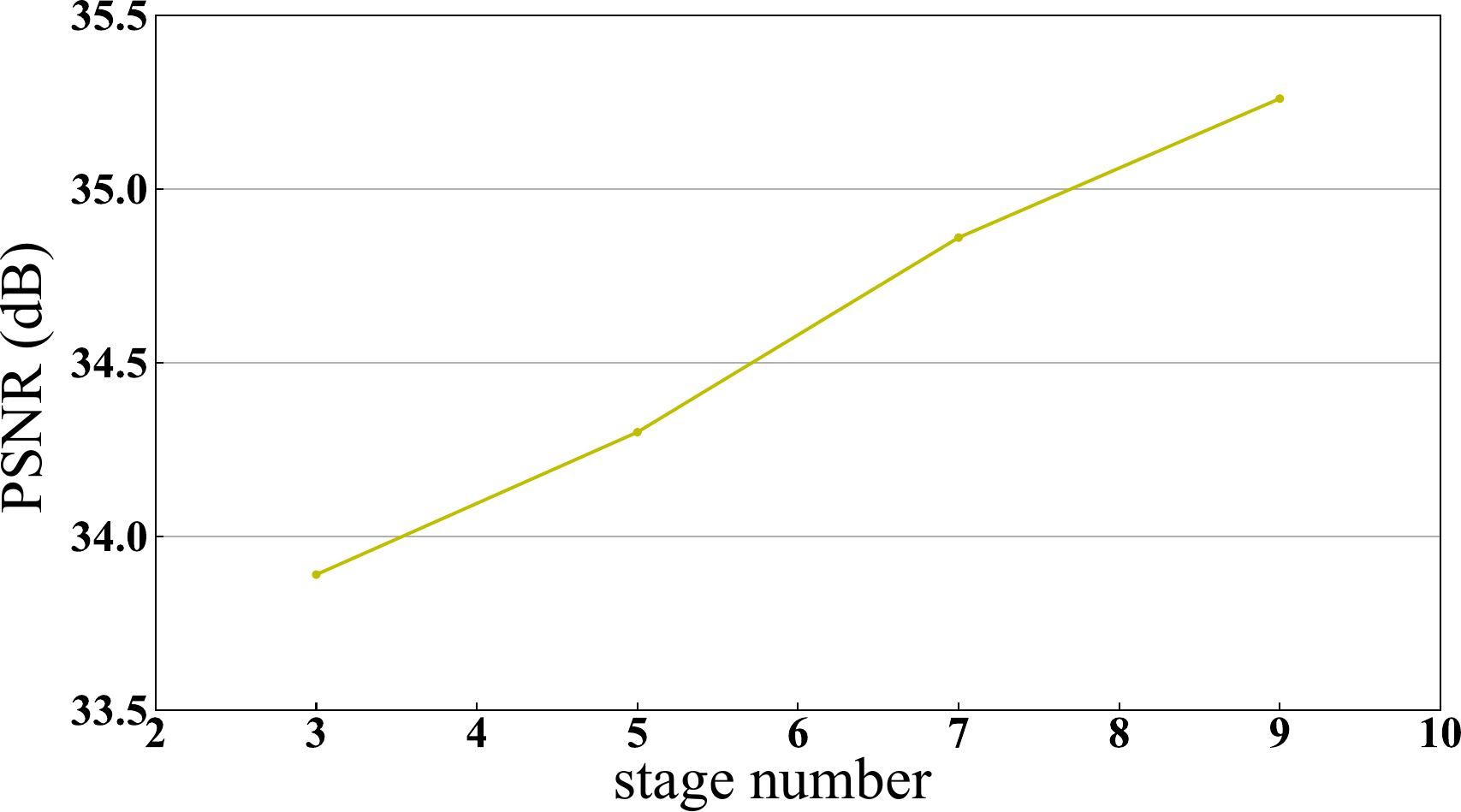}
    \vspace{-2mm}
    \caption{Effect of stage number on SCI reconstruction quality.}
    \vspace{-2mm}
    \label{fig:stage}
\end{figure}
\begin{figure}
    \centering
      \centerline{\includegraphics[width=0.9\linewidth]{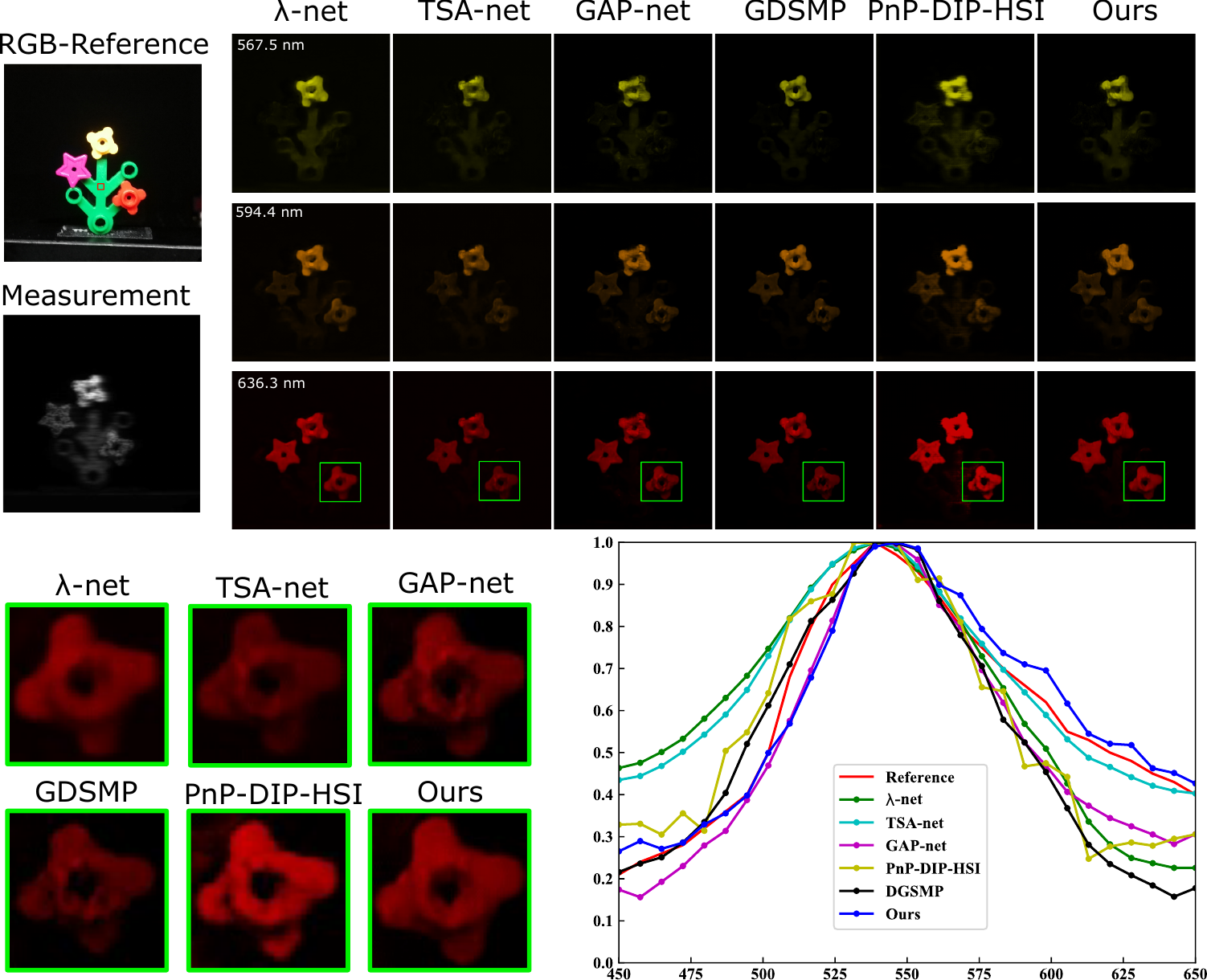}}
      \centerline{\includegraphics[width=0.9\linewidth]{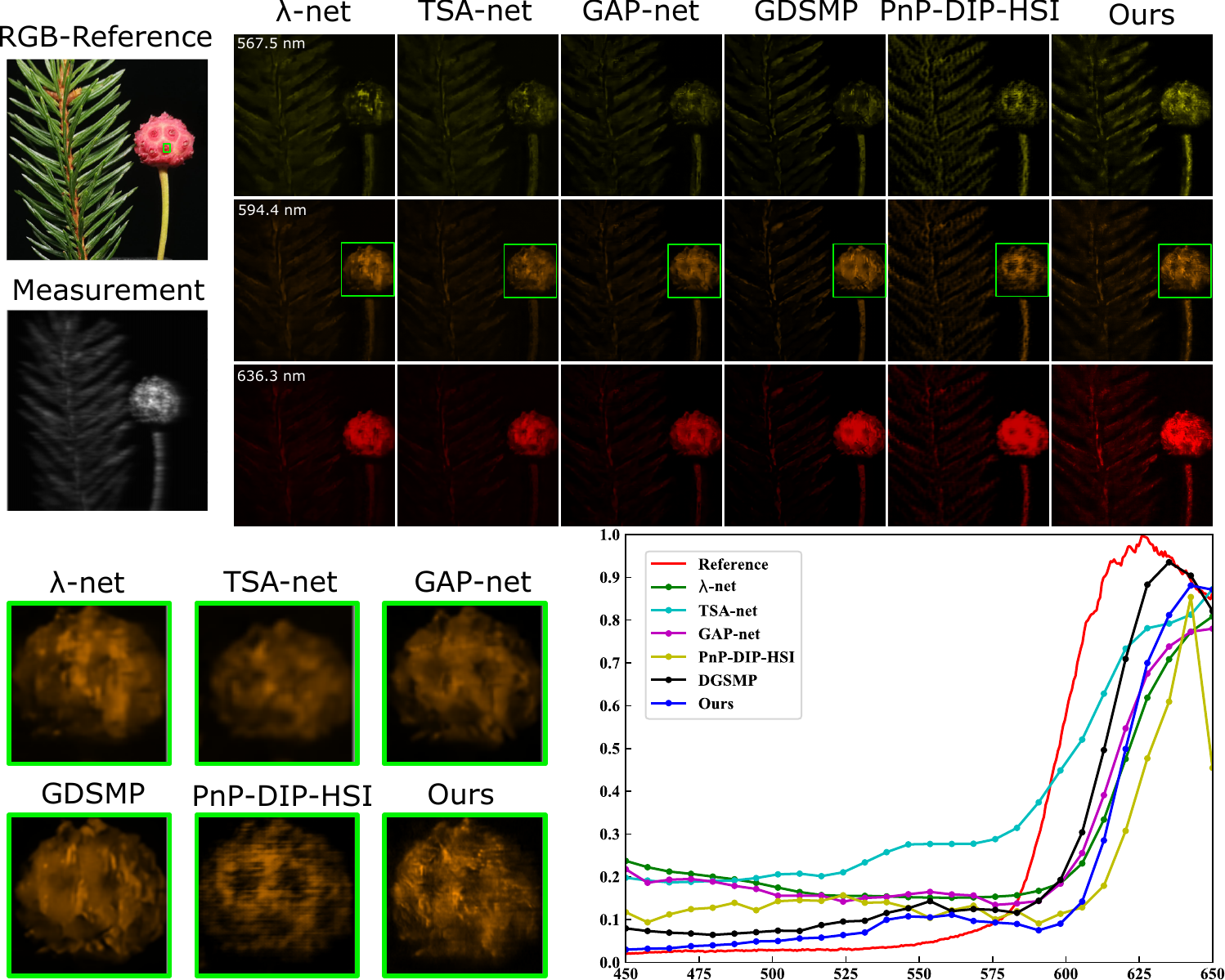}}
       \vspace{-2mm}
    \caption{Reconstruction results of GAP-CCoT and other spectral reconstruction algorithms 
  ($\lambda$-net, TSA-net, GAP-net, DGSMP, PnP-DIP-HSI) in  two real scenes (Scene 1 and Scene 2).}
  \label{fig:real_data}
   \vspace{-5mm}
\end{figure}

In order to verify the effectiveness of the contextual transformer and the GAP algorithm, 
we trained two different GAP-CCoT networks and two different Stacked CCoT networks (shown in  Fig.~\ref{fig:stack_cot}) for spectral SCI reconstruction respectively. 
Table~\ref{Tab:ablation} shows the reconstruction results of the two different networks we proposed, where
`w/o' CoT means removing the CoT branch at each stage of coding. 
We can clearly observe that the GAP-CCoT network is 0.99 dB higher in PSNR than
the Stacked CCoT network. The PSNR value of the CoT module is improved by 1.13 dB 
and 1.41 dB on the GAP-CCoT network and the Stacked CCoT network respectively.

In order to verify the impact of the number of stages on the reconstruction quality, we trained multiple models with different number of stages. As can be seen from Fig.~\ref{fig:stage}, the model proposed in this paper only needs three stages to complete high reconstruction quality, and the reconstruction quality increases with the increase of the number of stages.

\subsection{Real Data Results}
We test the proposed method on several real data captured by CASSI system \cite{Wagadarikar2008}. 
The system captures 28 spectral bands with wavelengths ranging from 450nm to 650nm. 
The spatial resolution of the object is $550\times{550}$, 
and the spatial resolution of the measurements captured by the plane sensor is $550\times{604}$. 
We compared our method with several SOTA methods 
($\lambda$-net \cite{Miao2019}, TSA-net \cite{Meng2020d}, 
GAP-net \cite{Meng2020b}, PnP-DIP-HSI \cite{Meng2021b}, 
DGSMP \cite{Huang2021b}) on real data. 
In addition to the results shown in Fig.~\ref{fig:best_results},  Fig.~\ref{fig:real_data} shows part of the visualization results and spectral curves of the reconstructed real data of another scene. 
By zooming in on a local area, 
we can see that our proposed method can recover more details and fewer artifacts. 
In addition, from the spectral correlation curve, our proposed method also has higher spectral accuracy.


\begin{table}[!htbp]
  \caption{Extending our method for {\bf Video Compressive Sensing}: The average PSNR in  dB, SSIM and running time per measurement of different algorithms on 6 benchmark datasets. 
  }
  \centering
  {
  \centering
  \begin{tabular}{c|c|c}
  \hline
  Algorithm
  & PSNR, SSIM 
  & Running time(s) 
  \\
  \hline
  \hline
  GAP-TV\cite{Yuan2016}
  & 26.73, 0.858
  & 4.201 (CPU) \\
  \hline
  PnP-FFDNet\cite{Yuan2020c} 
  & 29.70, 0.892
  & 3.010 (GPU)
  \\
  \hline
  DeSCI\cite{Liu2018}
  & 32.65, 0.935
  & 6180 (CPU)
  \\
  \hline
  BIRNAT\cite{Cheng2020b} 

  & 33.31, 0.951
  & 0.165 (GPU)
  \\
  \hline
  U-net\cite{Qiao2020} 
  & 29.45, 0.882
  & 0.031 (GPU)
  \\
  \hline 
  GAP-net-Unet-S12 \cite{Meng2020b} 
  & 32.86, 0.947
  & 0.007 (GPU)
  \\
  \hline 
  MetaSCI \cite{Wang2021e} 
  & 31.72, 0.926
  & 0.025 (GPU)
  \\
  \hline
  RevSCI \cite{Cheng2021d}
  & 33.92, 0.956
  & 0.190 (GPU)
  \\
  \hline
  \rowcolor{lightgray}
  Ours

  & 33.53, 0.954
  & 0.064 (GPU)
  \\
  \hline 
  \end{tabular}
  }
  \label{Tab:video}
\end{table}

\begin{figure}
  \centering 
  \includegraphics[width=1.0\linewidth]{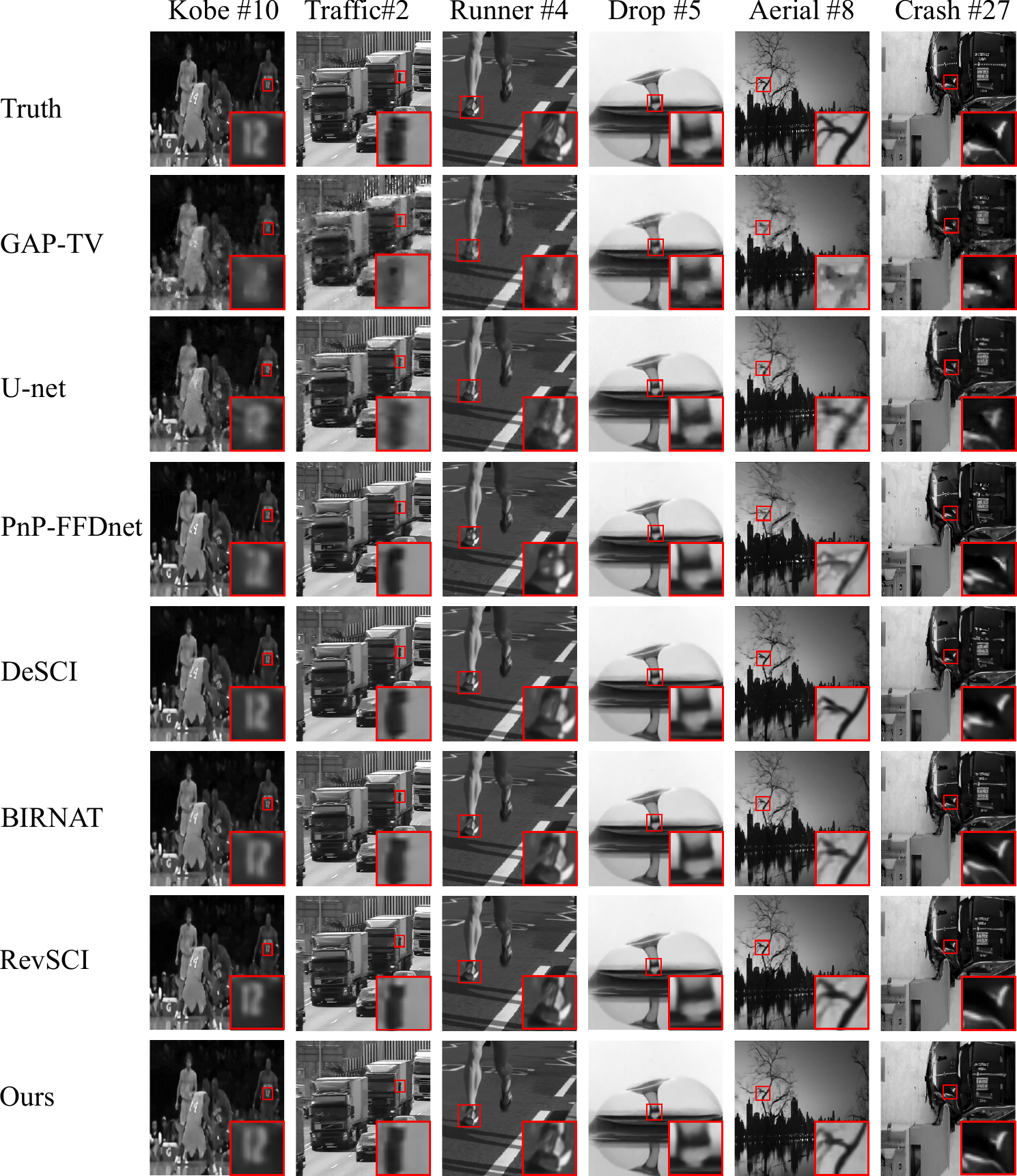}
  \caption{Reconstructed frame of our method and other algorithms (GAP-TV, DeSCI, PnP-FFDNet, U-net, BIRNAT, RevSCI) on 6 benchmark datasets.}
  \label{fig:vsci_sim}
\end{figure}
\section{Conclusions and Discussion}
In this paper, 
we use the inductive bias ability of convolution and 
the powerful modeling ability of transformer to propose a parallel module named CCoT, 
which can obtain more effective spectral features.
We integrate this module with the deep unfolding idea and the GAP algorithm, 
which can be well applied to SCI reconstruction. 
In addition, we have also developed similar models 
for video  compressive sensing \cite{Llull2013,Yuan2021a}
and our model leads to excellent results,
summarized in Table~\ref{Tab:video} and Fig.~\ref{fig:vsci_sim}. 
We believe that by fine-tuning the proposed networks, 
we should be able to achieve state-of-the-art results 
for video compressive sensing and also other reconstruction tasks.

{\small
\bibliographystyle{ieee_fullname}
\bibliography{scit_wangls}
}
\end{document}